\documentclass[10pt,A4]{article}
\usepackage{graphicx}
\usepackage{hyperref}
\usepackage{tikz}
\def\R2Lurl#1#2{\mbox{\href{#1}{\tt #2}}}

 \usepackage[]{hyperref}
\usepackage{amsmath, amsthm, amssymb}

\renewcommand{\qed}{\nobreak \ifvmode \relax \else
      \ifdim\lastskip<1.5em \hskip-\lastskip
      \hskip1.5em plus0em minus0.5em \fi \nobreak
      \vrule height0.75em width0.5em depth0.25em\fi}

\newcommand{\figi}{\begin{center}
\begin{tikzpicture}
\begin{scope}[scale=1.2]
\draw[color=black,ultra thin] (-4,-2)--(-4,2)--(4,2)--(4,-2)--(-4,-2); 
\draw[color=blue,densely dashed] (2.5,0)--(3,0);
\draw[color=blue,densely dashed] (2,0.5)--(2.0,1);
\draw[color=gray,ultra thick] (1.5,-0.5)--(1.5,0.5)--(2.5,0.5)--(2.5,-0.5)--(1.5,-0.5)--(1.5,0.5);
\draw[color=gray,thick] (1.6,-0.4)--(2.4,0.4);
\draw (2,1.2) node {{\footnotesize $\textup{\textbf{R}}=0$}};
\draw (3.4,0) node {{\footnotesize $\textup{\textbf{R}}=1$}};
\draw[color=blue,thick, ->] (0.1,0)--(1.5,0); 
\draw (0.75,-0.2) node {{\scriptsize $\tau_R$}};
\draw (0.75,0.2) node {{\scriptsize Beam}};
\draw (2,-1) node {{\scriptsize\parbox{2cm}{Polarizing cube\\ set at angle $\sigma_R$}}};
\draw[color=blue,densely dashed] (-2.5,0)--(-3,0);
\draw[color=blue,densely dashed] (-2,0.5)--(-2.0,1);
\draw[color=gray,ultra thick] (-1.5,-0.5)--(-1.5,0.5)--(-2.5,0.5)--(-2.5,-0.5)--(-1.5,-0.5)--(-1.5,0.5);
\draw[color=gray,thick] (-1.6,-0.4)--(-2.4,0.4);
\draw (-2,1.2) node {{\footnotesize $\textup{\textbf{L}}=0$}};
\draw (-3.4,0) node {{\footnotesize $\textup{\textbf{L}}=1$}};
\draw[color=blue,thick, ->] (-1.5,0)--(-0.1,0); 
\draw (-0.75,-0.2) node {{\scriptsize $\tau_L$}};
\draw (-0.75,0.2) node {{\scriptsize Photon}};
\draw (-2,-1) node {{\scriptsize\parbox{2cm}{Polarizing cube\\ set at angle $\sigma_L$}}};
\shade[inner color=yellow!50!white!50, outer color=yellow!50!white!50] (0,-1.5) -- (-3.8,-1.5) -- (-3.8,1.5) -- (0,1.5);
\draw (0.,-2.5) node {{\small \parbox{8cm}{\begin{center}
\emph{Figure 1: The classical set up -- right.}
\end{center}
}}};
\end{scope}
\end{tikzpicture}
\end{center}}

\newcommand{\figii}{\begin{center}
\begin{tikzpicture}
\begin{scope}[scale=1.2]

\draw[color=black,ultra thin] (-4,-2)--(-4,2)--(4,2)--(4,-2)--(-4,-2); 

\draw[color=blue,densely dashed] (2.5,0)--(3,0);
\draw[color=blue,densely dashed] (2,0.5)--(2.0,1);
\draw[color=gray,ultra thick] (1.5,-0.5)--(1.5,0.5)--(2.5,0.5)--(2.5,-0.5)--(1.5,-0.5)--(1.5,0.5);
\draw[color=gray,thick] (1.6,-0.4)--(2.4,0.4);
\draw (2,1.2) node {{\footnotesize $\textup{\textbf{R}}=0$}};
\draw (3.4,0) node {{\footnotesize $\textup{\textbf{R}}=1$}};
\draw[color=blue,thick, ->] (0.1,0)--(1.5,0); 
\draw (0.75,-0.2) node {{\scriptsize $\tau_R$}};
\draw (0.75,0.2) node {{\scriptsize Photon}};
\draw (2,-1) node {{\scriptsize\parbox{2cm}{Polarizing cube\\ set at angle $\sigma_R$}}};

\shade[inner color=yellow!50!white!50, outer color=yellow!50!white!50] (0,-1.5) -- (3.8,-1.5) -- (3.8,1.5) -- (0,1.5);

\draw[color=blue,densely dashed] (-2.5,0)--(-3,0);
\draw[color=blue,densely dashed] (-2,0.5)--(-2.0,1);
\draw[color=gray,ultra thick] (-1.5,-0.5)--(-1.5,0.5)--(-2.5,0.5)--(-2.5,-0.5)--(-1.5,-0.5)--(-1.5,0.5);
\draw[color=gray,thick] (-1.6,-0.4)--(-2.4,0.4);
\draw (-2,1.2) node {{\footnotesize $\textup{\textbf{L}}=0$}};
\draw (-3.4,0) node {{\footnotesize $\textup{\textbf{L}}=1$}};
\draw[color=blue,thick, ->] (-1.5,0)--(-0.1,0); 
\draw (-0.75,-0.2) node {{\scriptsize $\tau_L$}};
\draw (-0.75,0.2) node {{\scriptsize Beam}};
\draw (-2,-1) node {{\scriptsize\parbox{2cm}{Polarizing cube\\ set at angle $\sigma_L$}}};

\draw (0.,-2.5) node {{\small \parbox{8cm}{\begin{center}
\emph{Figure 2: The classical set up -- left.}
\end{center}
}}};
\end{scope}

\end{tikzpicture}
\end{center}}

\newcommand{\figiii}{\begin{center}
\begin{tikzpicture}
\begin{scope}[scale=1.2]

\draw[color=black,ultra thin] (-4,-2)--(-4,2)--(4,2)--(4,-2)--(-4,-2); 

\draw[color=blue,densely dashed] (2.5,0)--(3,0);
\draw[color=blue,densely dashed] (2,0.5)--(2.0,1);
\draw[color=gray,ultra thick] (1.5,-0.5)--(1.5,0.5)--(2.5,0.5)--(2.5,-0.5)--(1.5,-0.5)--(1.5,0.5);
\draw[color=gray,thick] (1.6,-0.4)--(2.4,0.4);
\draw (2,1.2) node {{\footnotesize $\textup{\textbf{R}}=0$}};
\draw (3.4,0) node {{\footnotesize $\textup{\textbf{R}}=1$}};
\draw[color=blue,thick, ->] (0.1,0)--(1.5,0); 
\draw (0.75,-0.2) node {{\scriptsize $\tau_R$}};
\draw (0.75,0.2) node {{\scriptsize Photon}};
\draw (2,-1) node {{\scriptsize\parbox{2cm}{Polarizing cube\\ set at angle $\sigma_R$}}};

\draw[color=blue,densely dashed] (-2.5,0)--(-3,0);
\draw[color=blue,densely dashed] (-2,0.5)--(-2.0,1);
\draw[color=gray,ultra thick] (-1.5,-0.5)--(-1.5,0.5)--(-2.5,0.5)--(-2.5,-0.5)--(-1.5,-0.5)--(-1.5,0.5);
\draw[color=gray,thick] (-1.6,-0.4)--(-2.4,0.4);
\draw (-2,1.2) node {{\footnotesize $\textup{\textbf{L}}=0$}};
\draw (-3.4,0) node {{\footnotesize $\textup{\textbf{L}}=1$}};
\draw[color=blue,thick, ->] (-1.5,0)--(-0.1,0); 
\draw (-0.75,-0.2) node {{\scriptsize $\tau_L$}};
\draw (-0.75,0.2) node {{\scriptsize Photon}};
\draw (-2,-1) node {{\scriptsize\parbox{2cm}{Polarizing cube\\ set at angle $\sigma_L$}}};

\shade[inner color=yellow!50!white!50, outer color=yellow!50!white!50] (0,-1.5) -- (-3.8,-1.5) -- (-3.8,1.5) -- (0,1.5);

\draw (0.,-2.5) node {{\small \parbox{8cm}{\begin{center}
\emph{Figure 3: The QM set up -- right.}
\end{center}
}}};

\end{scope}

\end{tikzpicture}
\end{center}}

\newcommand{\figv}{\begin{center}
\begin{tikzpicture}
\begin{scope}[scale=1.2]

\draw[color=black,ultra thin] (-4,-2)--(-4,2)--(4,2)--(4,-2)--(-4,-2); 

\draw[color=blue,densely dashed] (2.5,0)--(3,0);
\draw[color=blue,densely dashed] (2,0.5)--(2.0,1);
\draw[color=gray,ultra thick] (1.5,-0.5)--(1.5,0.5)--(2.5,0.5)--(2.5,-0.5)--(1.5,-0.5)--(1.5,0.5);
\draw[color=gray,thick] (1.6,-0.4)--(2.4,0.4);
\draw (2,1.2) node {{\footnotesize $\textup{\textbf{R}}=0$}};
\draw (3.4,0) node {{\footnotesize $\textup{\textbf{R}}=1$}};
\draw (2,-1) node {{\scriptsize\parbox{2cm}{Polarizing cube\\ set at angle $\sigma_R$}}};

\draw[color=blue,densely dashed] (-2.5,0)--(-3,0);
\draw[color=blue,densely dashed] (-2,0.5)--(-2.0,1);
\draw[color=gray,ultra thick] (-1.5,-0.5)--(-1.5,0.5)--(-2.5,0.5)--(-2.5,-0.5)--(-1.5,-0.5)--(-1.5,0.5);
\draw[color=gray,thick] (-1.6,-0.4)--(-2.4,0.4);
\draw (-2,1.2) node {{\footnotesize $\textup{\textbf{L}}=0$}};
\draw (-3.4,0) node {{\footnotesize $\textup{\textbf{L}}=1$}};
\draw[color=blue,thick, ->] (-1.5,0)--(1.5,0); 
\draw (0,-0.2) node {{\scriptsize $\tau=\tau_L=\tau_R$}};
\draw (0,0.2) node {{\scriptsize Beam}};
\draw (-2,-1) node {{\scriptsize\parbox{2cm}{Polarizing cube\\ set at angle $\sigma_L$}}};

\draw (0.,-2.5) node {{\small \parbox{8cm}{\begin{center}
\emph{Figure 5: The full classical set up.}
\end{center}
}}};
\end{scope}

\end{tikzpicture}
\end{center}}

\newcommand{\figiv}{\begin{center}
\begin{tikzpicture}
\begin{scope}[scale=1.2]

\draw[color=black,ultra thin] (-4,-2)--(-4,2)--(4,2)--(4,-2)--(-4,-2); 

\draw[color=blue,densely dashed] (2.5,0)--(3,0);
\draw[color=blue,densely dashed] (2,0.5)--(2.0,1);
\draw[color=gray,ultra thick] (1.5,-0.5)--(1.5,0.5)--(2.5,0.5)--(2.5,-0.5)--(1.5,-0.5)--(1.5,0.5);
\draw[color=gray,thick] (1.6,-0.4)--(2.4,0.4);
\draw (2,1.2) node {{\footnotesize $\textup{\textbf{R}}=0$}};
\draw (3.4,0) node {{\footnotesize $\textup{\textbf{R}}=1$}};
\draw[color=blue,thick, ->] (0.1,0)--(1.5,0); 
\draw (0.75,-0.2) node {{\scriptsize $\tau_R$}};
\draw (0.75,0.2) node {{\scriptsize Photon}};
\draw (2,-1) node {{\scriptsize\parbox{2cm}{Polarizing cube\\ set at angle $\sigma_R$}}};

\shade[inner color=yellow!50!white!50, outer color=yellow!50!white!50] (0,-1.5) -- (3.8,-1.5) -- (3.8,1.5) -- (0,1.5);

\draw[color=blue,densely dashed] (-2.5,0)--(-3,0);
\draw[color=blue,densely dashed] (-2,0.5)--(-2.0,1);
\draw[color=gray,ultra thick] (-1.5,-0.5)--(-1.5,0.5)--(-2.5,0.5)--(-2.5,-0.5)--(-1.5,-0.5)--(-1.5,0.5);
\draw[color=gray,thick] (-1.6,-0.4)--(-2.4,0.4);
\draw (-2,1.2) node {{\footnotesize $\textup{\textbf{L}}=0$}};
\draw (-3.4,0) node {{\footnotesize $\textup{\textbf{L}}=1$}};
\draw[color=blue,thick, ->] (-1.5,0)--(-0.1,0); 
\draw (-0.75,-0.2) node {{\scriptsize $\tau_L$}};
\draw (-0.75,0.2) node {{\scriptsize Photon}};
\draw (-2,-1) node {{\scriptsize\parbox{2cm}{Polarizing cube\\ set at angle $\sigma_L$}}};

\draw (0.,-2.5) node {{\small \parbox{8cm}{\begin{center}
\emph{Figure 4: The QM set up -- left.}
\end{center}
}}};
\end{scope}

\end{tikzpicture}
\end{center}}

\newcommand{\figvi}{\begin{center}
\begin{tikzpicture}
\begin{scope}[scale=1.2]

\draw[color=black,ultra thin] (-4,-2)--(-4,2)--(4,2)--(4,-2)--(-4,-2); 

\draw[color=blue,densely dashed] (2.5,0)--(3,0);
\draw[color=blue,densely dashed] (2,0.5)--(2.0,1);
\draw[color=gray,ultra thick] (1.5,-0.5)--(1.5,0.5)--(2.5,0.5)--(2.5,-0.5)--(1.5,-0.5)--(1.5,0.5);
\draw[color=gray,thick] (1.6,-0.4)--(2.4,0.4);
\draw (2,1.2) node {{\footnotesize $\textup{\textbf{R}}=0$}};
\draw (3.4,0) node {{\footnotesize $\textup{\textbf{R}}=1$}};
\draw (2,-1) node {{\scriptsize\parbox{2cm}{Polarizing cube\\ set at angle $\sigma_R$}}};

\draw[color=blue,densely dashed] (-2.5,0)--(-3,0);
\draw[color=blue,densely dashed] (-2,0.5)--(-2.0,1);
\draw[color=gray,ultra thick] (-1.5,-0.5)--(-1.5,0.5)--(-2.5,0.5)--(-2.5,-0.5)--(-1.5,-0.5)--(-1.5,0.5);
\draw[color=gray,thick] (-1.6,-0.4)--(-2.4,0.4);
\draw (-2,1.2) node {{\footnotesize $\textup{\textbf{L}}=0$}};
\draw (-3.4,0) node {{\footnotesize $\textup{\textbf{L}}=1$}};
\draw[color=blue,thick, ->] (-1.5,0)--(1.5,0); 
\draw (0,-0.2) node {{\scriptsize $\tau_L, \tau_R$}};
\draw (0,0.2) node {{\scriptsize Photon}};
\draw (-2,-1) node {{\scriptsize\parbox{2cm}{Polarizing cube\\ set at angle $\sigma_L$}}};

\draw (0.,-2.5) node {{\small \parbox{8cm}{\begin{center}
\emph{Figure 6: The full quantum set up.}
\end{center}
}}};
\end{scope}

\end{tikzpicture}
\end{center}}

\title{Does Time-Symmetry Imply Retrocausality? How the Quantum World Says ``Maybe''}

\author{Huw Price\thanks{Address: Centre for Time, Main Quadrangle A14, University of Sydney, NSW 2006, Australia. E-mail:  \href{mailto:hp331@cam.ac.uk}{hp331@cam.ac.uk}.
}
}

\date{October 19, 2011}

\begin{document}
\maketitle
\begin{abstract}
\noindent 
It has often been suggested that retrocausality offers a solution to some of the puzzles of quantum mechanics: e.g., that it allows a Lorentz-invariant explanation of Bell correlations, and other manifestations of quantum nonlocality, without action-at-a-distance. Some writers have argued that time-symmetry counts in favour of such a view, in the sense that retrocausality would be a natural consequence of a truly time-symmetric theory of the quantum world. Critics object that there is complete time-symmetry in classical physics, and yet no apparent retrocausality. Why should the quantum world be any different?

This note throws some new light on these matters. I call attention to a respect in which quantum mechanics is different, under some assumptions about quantum ontology. Under these assumptions,  the combination of time-symmetry without retrocausality  is unavailable in quantum mechanics, for reasons intimately connected with the differences between classical and quantum physics (especially the role of discreteness in the latter). Not all interpretations of quantum mechanics share these assumptions, however, and in those that do not, time-symmetry does not entail retrocausality.\\\ 

\end{abstract}

\section{Introduction}

Many writers have proposed that some puzzles of quantum mechanics (QM) might find elegant resolution, if we allow that there is retrocausality in the quantum world. Perhaps most importantly, retrocausality  has been claimed to offer an attractive path to a Lorentz-invariant explanation of Bell correlations, and other manifestations of quantum nonlocality, without action-at-a-distance (see, e.g.,  [\ref{ref:costa}]--[\ref{ref:cramer86}],  [\ref{ref:hokk}], [\ref{ref:miller96}]--[\ref{ref:price96}], [\ref{ref:suth83}]--[\ref{ref:wharton09}]). In general, it is a central assumption of Bell's Theorem~[\ref{ref:bell}] and other No Hidden Variable theorems in QM that there is no retrocausality: that hidden variables (HVs) are independent of future measurement settings. Prima facie, then, admitting retrocausality might be expected to make a big difference to the prospects for HV approaches to QM -- i.e., for the view that QM provides only an \emph{incomplete} description of the quantum world.

Some advocates of retrocausality in QM (see, e.g.,  [\ref{ref:costa76}], [\ref{ref:costa77}], [\ref{ref:miller96}], [\ref{ref:price97}]) have suggested  that considerations of time-symmetry count in its favour:  that if QM were made ``more time-symmetric,'' retrocausality would be a natural consequence. Against this, critics object that there is complete time-symmetry in classical physics, and yet no apparent retrocausality. Why should QM be any different?\footnote{I don't know an example of this objection in print, but I have often heard it in person.}

The main purpose of this note is to call attention to a respect in which QM is demonstrably different, under some assumptions about quantum ontology. Under these assumptions, a simple (and, so far as I know, previously unnoticed) argument shows that a time-symmetric ontology necessarily  puts the properties of a system 
\emph{prior to a measurement} under the control of an experimenter who chooses a setting for the measurement in question. Moreover, the argument turns on a feature of QM always thought to constitute one of the fundamental differences between classical and quantum physics, viz., the role in the latter of the new kind of discreteness introduced by quantisation.

However, the assumptions required by the argument are far from uncontroversial, and are easily seen to fail in a range of familiar interpretations of QM. So the argument does not support the sweeping claim that time-symmetry in QM requires retrocausality. But it does suggest that retrocausality is ``close to the surface'' in QM, in the sense it is a direct consequence of time-symmetry, under assumptions which many writers (past and present) would regard as well-motivated, even if not indisputable.  

Roughly speaking, the interpretations  in which the assumptions fail are of two kinds: (i) instrumentalist views of QM, which reject the kind of ontology ``between measurements'' which is needed to provide a domain of retro-effects; and (ii) all no-collapse versions of ontic realism about the wave function, which remove the particular kind of discreteness on which the argument depends.

Leaving instrumentalism to one side, the crucial issue is whether the required discreteness of measurement outcomes is regarded as ``all there is,'' or whether continuity is provided elsewhere in the ontology of one's model, e.g., in the form of an ontic wave function (without collapse). In the former case  -- i.e., roughly, if we are any sort of realist other than an Everettian or a Bohmian --  the argument shows that we do need to make a choice between time-asymmetric ontology and retrocausality in QM, in a manner not true of classical physics.\footnote{The situation for Everettian and Bohmian views is less clear. There may be other reasons to think that they cannot entirely combine time-symmetry and one-way causality. (I discuss the case of the  Bohm view in \S7 below.) But they escape a particularly sharp argument for the incompatibility of these options which applies to other realist views.}
In this case, the usual assumption  that HVs are independent of future measurement settings is incompatible with time-symmetry. 
\figi\vspace{12pt}
\figii\vspace{6pt}

\section{Polarization: classical and quantum}

\subsection{The classical case}

We first consider the standard description in classical electromagnetism (CEM) of the apparatus shown in Figure 1. A beam of light, linearly polarized in direction $\tau_R$, is directed towards an ideal polarizing cube set at angle $\sigma_R$. CEM predicts that the beam will split into two output beams. The \emph{transmission} beam, here labelled $\textup{\textbf{R}}=1$, will have an intensity $\cos^{2}(\tau_R - \sigma_R)$ times that of the input beam. The \emph{reflection} beam, here labelled $\textup{\textbf{R}}=0$, will have an intensity {$\sin^{2}(\tau_R - \sigma_R)$ times that of the input beam. Thus in the case in which $\tau_R = \sigma_R$, all the energy goes on the transmission beam; in the case in which $\tau_R = \sigma_R + \pi/2$, all the energy goes on the reflection beam; and other cases are distributed between these extremes, in a continuous fashion. The transmission ($\textup{\textbf{R}}=1$) and reflection ($\textup{\textbf{R}}=0$) beams have linear polarization in directions  $\sigma_R$ and  $\sigma_R + \pi/2$, respectively.

Figure 1 may be taken to represent a class of possible experiments, as $\tau_R$ and $ \sigma_R$ are allowed to vary. It follows from the time-symmetry of CEM that for any instance of this class there is a corresponding ``reverse'' experiment, in the intuitive sense of ``reverse'' in which moving from A to B is the reverse of moving from B to A. In an experiment of this reversed form, two beams of linearly polarized light, with appropriate polarization angles, intensities and phases, are combined into a \emph{single} beam, on passage through a polarizing cube. 

Figure 2 depicts a typical case of this reversed kind. The diagram has been mirror-reversed compared to Figure 1 (thus preserving the convention that inputs come from the left and outputs go to the right), in order to make it easy (below)  to depict the case in which the \emph{output} from an experiment of this reversed kind is used as \emph{input}  for a case of the first kind. The mathematical constraints are exactly as one would expect from time-symmetry: if the input on the transmission channel, $\textup{\textbf{L}}=1$, is polarized in direction $\sigma_L$ with intensity $\cos^{2}(\tau_L - \sigma_L)$, the input on the reflection channel,  $\textup{\textbf{L}}=0$, is polarized in direction $\sigma_L + \pi/2$ with intensity $\sin^{2}(\tau_L - \sigma_L)$, and the phases of the two beams are related exactly as in the original case, then there is an output beam with intensity $1$ and polarization $\tau_L$, in the direction of the input beam in the corresponding experiment of the original form.\footnote{If this seems surprising, keep the symmetry of the two classes of experiments in mind. Each instance of the new kind is simply the time-reverse of an instance of the old, in the intuitive sense mentioned above. If you were shown a video of such an experiment, with any dissipative effects suppressed, you would be unable to tell whether the video was being played forwards or backwards.}

\subsection{The quantum case}

Figures 3 and 4 show the quantum versions of these kinds of experiments, in the single photon case. For the moment, to highlight the feature which will be of interest, we assume that the outputs at $\textup{\textbf{R}}=1$ and $\textup{\textbf{R}}=0$ and the inputs at $\textup{\textbf{L}}=1$ and $\textup{\textbf{L}}=0$ are now  ``discrete,'' in the sense that the photon leaves or enters the apparatus on one channel or other. (More later on the case in which the input and/or output may be a superposition.) 

The factors $\cos^{2}(\tau_R - \sigma_R)$ and $\cos^{2}(\tau_L - \sigma_L)$ now represent probabilities, rather than intensities. 
In Figure 3, a photon with polarization $\tau_R$ has a probability $\cos^{2}(\tau_R - \sigma_R)$ of being detected on the  $\textup{\textbf{R}}=1$ channel, and a probability $\sin^{2}(\tau_R - \sigma_R)$ of being detected on the  $\textup{\textbf{R}}=0$ channel. (The intensity interpretation is recovered in the limit, as the numbers of photons goes to infinity.)

There are some subtleties about what these probabilities amount to, on the input side. Intuitively, we want to say that  in Figure 4, a photon emerging with polarization $\tau_L$ has a probability $\cos^{2}(\tau_L - \sigma_L)$ of \emph{having entered the apparatus} on the  $\textup{\textbf{L}}=1$ channel, and a probability $\sin^{2}(\tau_L - \sigma_L)$ of having entered on the  $\textup{\textbf{L}}=0$ channel. But such retrodictive probabilities are unreliable in many circumstances. If we know that the only photon source lies on the   
$\textup{\textbf{L}}=1$ channel, for example, this information would incline us to say that the probability was close to one that an emerging photon with polarization $\tau_L$ had come from that channel, independently of $\cos^{2}(\tau_L - \sigma_L)$ (at least unless $\cos^{2}(\tau_L - \sigma_L)\approx 0$).

We can deal with this issue by stipulating that we lack such information: e.g., by requiring that the choice of input channel is chosen at random, with equal probability. In these circumstances, the factors $\cos^{2}(\tau_L - \sigma_L)$  and  $\sin^{2}(\tau_L - \sigma_L)$ do supply reliable retrodictive probabilities -- specifically, the conditional probabilities of input on the $\textup{\textbf{R}}=1$ and $\textup{\textbf{R}}=0$ channels, respectively, given an output of a photon with polarization $\tau_L$, in the direction shown. (Again, the intensity interpretation is then recovered in the limit, as the numbers of photons goes to infinity.)\vspace{12pt}

\figiii\vspace{12pt}
\figiv

\figv

\section{Experimental control}

We now consider the question of what can be controlled, in the CEM and quantum cases, by experimenters who control \emph{only the polarizer settings,} $\sigma_L$ and $\sigma_R$. The reason for this restriction on what the experimenters control will become clear as we proceed.

\subsection{The classical case}

Figure 5 shows the CEM version of experiment we obtain by combining a pair of experiments of the two previous classes. What interests us is the question of what control, if any, the lefthand experimenter, Lena, has over the intermediate polarization, \emph{if she has no control over the inputs at} $\textup{\textbf{L}}=1$ \emph{and} $\textup{\textbf{L}}=0$.

To put this question in stark form, imagine that the inputs $\textup{\textbf{L}}=1$ and $\textup{\textbf{L}}=0$ are controlled by a Demon, who knows what setting $\sigma_L$ Lena has chosen for her polarizer. It is not difficult to see that under these conditions the Demon can produce any intermediate polarization $\tau$ he wishes, by an appropriate selection of inputs. This follows directly from the time-symmetry CEM: the inputs the Demon needs on the left are exactly the outputs Nature produces on the right, with intermediate polarization $\tau$ and right setting $\sigma_R=\sigma_L$.

Lena's lack of control of $\tau$ on the left, in the case in which the Demon controls the inputs, is exactly mirrored for her sister experimenter, Rena, on the right. The intuitive reason why Rena cannot control $\tau$ by varying the right polarizer setting $\sigma_R$ is that (as we just noted) Nature is able to make up for any difference in $\sigma_R$ by a difference in intensities of the output beams $\textup{\textbf{R}}=1$ and $\textup{\textbf{R}}=0$. In this case, $\tau$ doesn't shift, no matter how much Rena wiggles the setting $\sigma_R$. This is why there need be no retrocausality in this case, of course.\footnote{For future reference, note that it does not follow that there \emph{could not be} retrocausality. To block retrocausality altogether, Nature must behave as the mirror image of the perfect obstructive Demon. Imperfect Demons, who block some but not all forward influence on the left, correspond to ways in which Nature \emph{might be,} to allow some retrocausality on the right.}

To characterise this situation, I shall say that Lena has no \emph{input-independent} control of $\tau$, and that Rena has no \emph{output-independent} control of $\tau$. The importance of the notion of input-independence is that it mimics on the front end of the experiment precisely the question we need to consider on the back end, to think about the possibility of retrocausality. For on the back end, on the right, Rena has control of the \emph{setting,} $\sigma_R$, but not the \emph{outcome.} As we have just seen, the issue of retrocausality turns on the question as to what else Rena can influence, if her control is limited in this way. Input-independence enables us to think about the analogous issue at the front end of the apparatus, and time-symmetry enables us to move from one case to the other. 
The next step is to consider these issues in the quantum case.\vspace{12pt}

\figvi

\subsection{The quantum  case}

Figure 6 shows the analogous single photon case. (It will become clear in a moment why we need to distinguish between $\tau_L$ and $\tau_R$, in this case.)

For the moment, we are still assuming that the photon enters the apparatus on the left on one channel or other. (In other words, that the Demon cannot supply photons in a superposition of $\textup{\textbf{L}}=1$ and $\textup{\textbf{L}}=0$.) It is a striking fact that in this  case, Lena has more control over the left-polarization $\tau_L$ than  in the classical case, even if the Demon retains control of the inputs at $\textup{\textbf{L}}=1$ and $\textup{\textbf{L}}=0$. Of course, the Demon can refuse to provide any photons at all, in which case nothing emerges from the Lena's apparatus. But if we assume that the Demon shares with Lena the goal of emitting photons to the right, Lena now has complete control of  $\tau_L$,  $\textup{mod\ }\pi/2$. For if Lena sets her polarizer to angle $\sigma_L=\tau_L$, the Demon's only choices are to send in a photon via $\textup{\textbf{L}}=1$, in which case it emerges with polarization $\tau_L$; or to send in a photon via $\textup{\textbf{L}}=0$, in which case it emerges with polarization $\tau_L + \pi/2$.\footnote{The physical direction of the $\textup{\textbf{L}}=0$ beam depends on $\sigma_L$, but this creates no additional difficulty for the Demon, under our assumption that he knows what setting Lena has chosen.}

Thus in the notation introduced above, Lena now has \emph{input-independent} control of $\tau_L$, up to an additive factor of $\pi/2$, by means of her control of the left setting, $\sigma_L$. Note that this additional degree of control on Lena's part results directly from restriction on the Demon's options imposed by the assumption the input photon be on one channel or other. In effect, Lena is attempting to control a continuously-valued output variable ($\tau_L$) by manipulating a continuously-valued input variable ($\sigma_L$). The Demon can counteract any such influence, if he has an appropriate continuously-valued input variable of his own to play with, as in the classical case above. But if the Demon is restricted to a finite, \emph{discrete} range of inputs -- $n,$ say, in the general case -- then the best he can do is to group the possible outputs into $n$-tuples, in such a way that Lena cannot control which member of a particular $n$-tuple the output turns out to be; but can control from which $n$-tuple the choice is to be made.

This restriction on the Demon's abilities, and a corresponding restriction on Nature at the other end of the experiment, will play a crucial role in what follows; I shall call it \textbf{Discreteness}. I want to emphasise two points about Discreteness, at this stage. The first is that although it is an idea which -- at least naively -- seems at home in the quantum context, it doesn't depend on the quantum framework: we could  impose Discreteness in the classical case, too, by stipulating that the Demon can only use one input channel at a time. The second is that I am not claiming that Discreteness is a realistic restriction, even in the quantum case: on the contrary, as we shall see, we can restore continuity in the quantum case, by allowing the Demon to input superpositions of $\textup{\textbf{L}}=1$ and $\textup{\textbf{L}}=0$.
The real interest of Discreteness in the quantum case stems from the fact that it does turn out to be a realistic restriction, at least under some views of quantum ontology, at the other end of the experiment.\footnote{To avoid a possible confusion, let me also stress that my use of the terms \emph{discrete} and \emph{Discreteness} is not derived from the familiar distinction between discrete and continuous spectra, in relation to quantum observables. The observables we shall be discussing are certainly discrete (in fact, $2$-valued) in the latter sense. As we shall see, however, this is compatible, in some views of quantum ontology, with the outputs not being discrete in the sense relevant to my argument.} 

We have seen that Discreteness gives Lena some input-independent control of $\tau_L$, on the left side of our experiment. 
What does it entail on the other side of the experiment? In the classical case, we noted that Lena's lack of  input-independent control of $\tau$ is mirrored for Rena, on the right. Rena has no output-independent control of $\tau$, so long as Nature absorbs any change she (Rena) makes to the setting $\sigma_R$ in changes in the output intensities (thereby requiring no change in $\tau$ itself). As we noted, this is why there need be no retrocausality in the classical case (despite complete time-symmetry).

To ask the corresponding question in the quantum case, it will be helpful to make explicit another assumption, which I shall call \textbf{Realism}. It is that $\tau_L$ is a \emph{beable,} in John Bell's sense. As Bell (\ref{ref:bell2}, p.~174) puts it:
``The beables of the theory are those elements which might correspond to elements of reality, to things which exist.'' 
Thus Realism is the assumption that $\tau_L$ labels something \emph{real} -- some aspect of the \emph{ontology} of the quantum world, as philosophers would say. Some writers reject this assumption, in the quantum domain, at least in the sense that they think that the quantum state should not be understood as playing this role -- more on this option in \S5.1 below -- but for now, I assume it.

With Realism in place, we can now formulate a further assumption, viz., \textbf{Time-symmetry}. It stipulates that the ontology of the system depicted in Figure 6 be time-symmetric, in the intuitive sense introduced above: if a process is permitted by the theory, then so, equally, is the process we describe by replacing $t$ by $-t$ in a description of the original process -- by ``playing the video in reverse,'' in the familar analogy. 

In the quantum version of our experiment, Time-symmetry thus requires that there be an element of reality, $\tau_R$, that stands to $\sigma_R$, $\textup{\textbf{R}}=1$ and $\textup{\textbf{R}}=0$ in precisely the way that $\tau_L$ stands to $\sigma_L$, $\textup{\textbf{L}}=1$ and $\textup{\textbf{L}}=0$. Why? Because if there were not, the absence of  such a $\tau_R$ in a complete description of the experiment would easily allow us to tell whether a video of a particular run of the experiment was being shown to us ``forwards'' or ``backwards.'' When it is shown forwards, we see $\tau_L$ correlated with $\sigma_L$, $\textup{\textbf{L}}=1$ and $\textup{\textbf{L}}=0$ on the left; when it is shown backwards, we appear to see  $\tau_R$ correlated with $\sigma_R$, $\textup{\textbf{R}}=1$ and $\textup{\textbf{R}}=0$ on the right. If Time-symmetry does not hold -- and there is no such $\tau_R$, in fact -- then the latter case cannot be veridical, but must result from reversal of the movie.\footnote{It is not a relevant objection that no video could actually be made of such processes, due to the obvious limits on observability. We simply need to imagine a computer-generated movie, depicting the evolution of the assumed ontology.} 

Note that the textbook model of QM, which combines evolution of the state function in accordance with Schr\"odinger's Equation with collapse in accordance with the Projection Postulate, is manifestly \emph{not} time-symmetric, in this sense. As is well-known, it is the Projection Postulate that introduces a time-asymmetry. (It requires that the state be an eigenvector of the operator corresponding to an observable \emph{after} but not \emph{before} the measurement in question.) I shall say a little more about the connection between this feature of the textbook picture and our present concerns in \S5.3 below, but for the moment, it means that the standard picture is simply off the map, for present purposes -- of no direct relevance to the  issue as to whether if we \emph{do} assume Time-symmetry in QM, we are committed to retrocausality.

The assumptions Realism and Time-symmetry thus together make sense of the presence of the term $\tau_R$ in Figure 6, and entitle us to regard it as a beable, denoting an element of reality in just the way that  (Realism entails that) $\tau_L$ does. However, we saw that if we assume Discreteness on the left, Lena has some input-independent control over  $\tau_L$, by manipulating $\sigma_L$. If we now assume Discreteness on the right -- i.e., if we assume that a single photon emerges either at $\textup{\textbf{R}}=1$ or at $\textup{\textbf{R}}=0$ -- then it follows by symmetry that Rena has precisely the same output-independent control over the value of $\tau_R$ as her sister has over the value of $\tau_L$. She, too, can determine its value up to an additive factor of $\pi/2$, no matter what Nature does with the outputs on $\textup{\textbf{R}}=1$ and $\textup{\textbf{R}}=0$. 
In particular, this means that if Rena changes the setting $\sigma_R$ by any amount $\rho\neq\pi/2$, this will result in a \emph{different} value of $\tau_R$ for any subsequent photons. In counterfactual terms, it seems reasonable to say of any particular case that if she \emph{had} chosen $\sigma_R+\rho$ rather than $\sigma_R$, the value of $\tau_R$ would have been different. Intuitively, Rena thus has retrocausal control over $\tau_R$, $\textup{mod\ }\pi/2$.

As we wanted to show, then, time-symmetry does require retrocausality in this case, for reasons not present in CEM.\footnote{As before, we could duplicate the result in CEM by imposing the restriction that the output energy be entirely on one channel or other. However, without the flexibility provided by indeterminism in the QM case, a time-symmetric CEM model of this kind would lead to inconsistencies, in the sense that Lena and Rena could choose incompatible values of $\tau$. The distinction between $\tau_L$ and $\tau_R$ is needed to avoid this possibility in the QM case.} Once again,  the heavy lifting is done by Discreteness, for it is this that ensures that Nature lacks the range of variability required to absorb the difference of a change in 
$\sigma_R$ entirely in the future. Given the time-symmetric ontology required by the combination of Realism and Time-symmetry, Discreteness makes  retrocausality a simple consequence of the dynamical laws. 

Let us label this piece of reasoning the \emph{Time-Symmetry Argument for Retrocausality} (TSAR), and summarise its conclusions in the following form:
\begin{equation*}
{\textup{{Realism}}}+{\textup{Time-symmetry}}+{\textup{{Discreteness}}}\Rightarrow{\textup{{Retrocausality.}}}
\end{equation*}

\section{Discussion}

What is TSAR's significance? At a minimum, I think, the argument provides an interesting toy model, built from very familiar materials, that casts some useful light on the complex relationship between time-symmetry and retrocausality. The model begins with a case in which time-symmetry clearly does not require retrocausality; and then shows how a simple restriction on the space of possibilities available to Nature (in Her response to an agent's choices) can have the effect of imposing retrocausality, if time-symmetry is to be preserved. The model seems likely to throw some light, too, on the consistency constraints of a retrocausal framework -- as suggested above, the indeterminism in QM seems to play a crucial role at this point.

But does TSAR have any direct application to QM? In particular, can it be used to support the claim that time-symmetry in QM requires retrocausality? The toy model shows how time-symmetry \emph{can} require retrocausality, but is the world described by QM the kind of world in which this is true; or more like the world of CEM, in which it is not true? I want to conclude with some discussion of these questions. 

To focus the discussion, let me put on the table the following claim:

\begin{description}
\item[Symmetry-Requires-Retrocausality Thesis (SRRT):] TSAR establishes\\ that a time-symmetric ontology for photon polarization is necessarily retrocausal. Hence, if the quantum world is time-symmetric, it must be retrocausal.
\end{description}
I stress that I am not endorsing this thesis, but merely putting it on the table for discussion. I am interested in what counter-arguments are available to its opponents. 

The most powerful objections to SRRT are those that question the status of the Discreteness assumption on which TSAR relies, arguing that it is unrealistic, at least in some interpretations of QM. I turn to such arguments in \S6. Before that, however, I want briefly to mention a couple of responses which take a different tack, questioning the ontological assumptions of TSAR.

\section{Ontological options}

\subsection{Avoid the ontology altogether}
 The symmetry argument does not go through if we deny that the usual quantum polarization $\tau_L$ is a beable, or element of reality.  Hence SRRT can certainly be evaded by a sufficiently thoroughgoing instrumentalism about the quantum world. This loophole is not available to proponents of the Hidden Variable program, of course.\footnote{It is not even available  to \emph{opponents} of HVs, in so far as they are interested in the project of exhibiting supposedly undesirable consequences of the HV approach --  in this context, they cannot begin by denying their opponents' basic premise.}
 
 The costs of this loophole should not be underestimated, however. One way to see what some of these costs are is to note that the apparatus in Figure 6 can easily be used to signal, by an experimenter who controls both the input channel and the left polarizer setting $\sigma_L$. Under these circumstances,  $\tau_L$ would normally be thought to be what ``carries'' the signal, or causal influence, from one side of the apparatus to the other. Denying that there is any beable to play this role seems to amount to accepting a kind of action-at-a-distance (though timelike, rather than spacelike).\footnote{In [\ref{ref:evans}], Evans, Wharton and I argue that this option should be regarded as analogous to the standard view of spacelike nonlocality.}

\subsection{Avoid the specific ontology of the example}

A more subtle objection would be that the grounds for treating the usual quantum polarization $\tau_L$ as a beable rest on the assumption that there is no retrocausality. If the photon ``already knows''  $\sigma_R$, then it doesn't seem to need the full information carried by $\tau_L$, in order to explain the correlations we find in the full experiment, for variable $\sigma_L$ and $\sigma_R$ -- perhaps there is no such beable as $\tau_L$. 

One way to see this is to exploit the symmetry of the experiment. This symmetry entails that if the dependence of the correlation between  $\textup{\textbf{L}}$ and $\textup{\textbf{R}}$ on $\sigma_L$ and $\sigma_R$ can be explained (as in the orthodox picture) by postulating a beable $\tau_L$ to ``carry'' information from left to right, it could equally be explained by postulating a beable $\tau_R$ carrying information from right to left.\footnote{A reason for preferring one explanation to the other would have to come ``from outside'', for the symmetry implies that it cannot rest on anything internal to the experiment itself.} Since in the former case the explanation of the correlation does not require $\tau_R$, the explanation of the same correlation by the latter means  would not require $\tau_L$.
The latter explanation, by means of $\tau_R$, provides an extreme illustration of the general point: by making information about the ``future'' end of the experiment available at the ``past'' end, we thereby reduce the need for a beable to carry information from past to future. 

This is an important point, in my view. Among other things, it raises the question as to whether a time-symmetric model should involve the two beables $\tau_L$ and $\tau_R$, or rather some single beable, dependent both on $\sigma_L$ and $\sigma_R$ (which would combine the role of both $\tau_L$ and $\tau_R$, so to speak). On the face of it, models with two beables seem to fail to exploit the ontological efficiency possible in a retrocausal theory -- a fact which might be held to be a disadvantage of retrocausal models (e.g., those of [\ref{ref:aharonov}, \ref{ref:sutherland}, \ref{ref:wharton07}]) which proceed by adding an additional wave function (``coming from the future'') to standard QM, and interpreting both wave functions in an ontic manner.\footnote{Cramer's Transactional Interpretation  [\ref{ref:cramer86}]  seems to evade this objection, because its ``offer'' and ``confirmation'' waves do not have the same ontic status as the two wave functions in the models just cited.}

As an illustration of the ontological economy possible in a retrocausal theory, consider the following toy model. Replace the continuously-valued beables $\tau_L$ and $\tau_R$ with a single two-bit binary variable -- i.e., one of the four possibilities $\textbf{00}$, $\textbf{01}$, $\textbf{10}$ and $\textbf{11}$, where the first and second bits
represent the left (past) and right (future) input/output channels,
respectively.  Let the probability distribution over the four possible values of this two-bit variable depend on $\sigma_L$ and $\sigma_R$, in accordance with Malus' Law:
\begin{equation*}
\textup{Pr}(\textbf{00}\lor\textbf{11})=\cos^{2}(\sigma_L - \sigma_R)\textup{;\ }
\textup{Pr}(\textbf{01}\lor\textbf{10})=\sin^{2}(\sigma_L - \sigma_R).
\end{equation*}
This model reproduces the predictions of QM  (assuming Discreteness), and hence demonstrates  that if retrocausality is allowed, 
we need far less ontology for the intuitive causal
work of conveying information from one end of the experiment to the other. In
particular, we no longer need a beable to carry the continuous information
corresponding to the setting $\sigma_L$  -- in so far as it needs to be (given retrocausality), that variability can be entirely absorbed into the
statistics governing a discrete variable.\footnote{Indeed, we could economise even further, to a \emph{one-bit} variable representing the ``parity'' of the input--output channel comparison. If we map  $\textbf{00}$ and $\textbf{11}$ in the previous model to $\textbf{1}$, and $\textbf{01}$ and $\textbf{10}$ to $\textbf{0}$, and now stipulate that $\textup{Pr}(\textbf{0})=\cos^{2}(\sigma_L - \sigma_R)$, we again get the correct predictions. The difference is that in this simpler model the hidden variable ``tells'' the photon in the future whether it should take the same or a different output channel, compared to its input channel in the past, without telling it \emph{which} input channel it took in the past. The resulting correlations might then seem to involve  ``spooky action at a time-like distance,'' in comparison to those of the richer model.}

I think it is an interesting question whether the information structure and hence the ontological economy of this toy model could be replicated in a realistic physical model of the photon polarization case, or indeed of more complex cases. For the moment, however, I simply want to emphasise the relevance of this possibility to the present argument. On the one hand, it does reveal a significant lacuna in TSAR, as I presented it above, at least in the sense that it makes Realism seem ill-formulated: someone attracted to the approach described here will claim to be a realist by disposition, in the sense that they take seriously the question as to what exists ``between measurements,'' even though they do {not} -- as our version of Realism requires -- assume that $\tau_L$ labels an element of reality. On the other hand,  however, this point carries no weight at all as an objection to SRRT. An opponent of SRRT who sought to rely on it would simply shoot herself in the foot, invoking retrocausality in order to block an argument in favour of retrocausality. 

\subsection{Make the ontology time-asymmetric}
Another way to escape the consequences of SRRT -- i.e., to avoid a commitment to retrocausality -- is  to make one's ontology time-asymmetric. It is important to be clear  that this option is not an \emph{objection} to SRRT. SRRT merely claims that TSAR establishes that \emph{if} the ontology of QM is time-symmetric, then it is retrocausal. In other words, SRRT does not deny that we can avoid retrocausality by abandoning time-symmetry. It simply maintains that we need to make a choice between the two. This option concedes that we need to make a choice, but chooses time-asymmetry in preference to retrocausality.

As I remarked earlier (\S3.2), the textbook version of QM, with the Projection Postulate, is an example of a model that exploits this option. I want to make two notes in passing at this point. First, it is important to be clear that avoiding an argument requiring retrocausality is not the same thing as demonstrating that there is no retrocausality. I shall explore this distinction below with reference to  the de Broglie-Bohm model, but -- as I shall note there (\S7.1) -- the same point applies to the textbook picture: its case for thinking that the quantum world is not retrocausal seems remarkably weak.

Secondly, our discussion above hints at an additional cost of the textbook model, on top of the cost (if cost it be) of time-asymmetry itself. In \S5.2 we exploited the fact a model with a time-asymmetric boundary-independent ontology (i.e., in that case, with a beable $\tau_L$ to carry information from past to future) could easily be transformed into a different model, the time-reverse of the first, which would do an equally good job of explaining the correlations observed on the experiment in question. Hence the threat: unless there is  some independent reason to prefer one model to the other -- which would be, \emph{ipso facto,} a reason to expect time-asymmetry in the domain in question -- then the time-asymmetry will have introduced an undetectable fact of the matter, about which of two kinds of worlds we actually inhabit. (In the terminology of \S5.2, is it a $\tau_L$ world or a $\tau_R$ world?) We break the symmetry at the cost of making it undetectable which way it is broken. A time-symmetric model avoids this consequence.

\subsection{Conclusion}

The prospects for rejecting SRRT on ontological grounds are slim. TSAR can be avoided by a thoroughgoing instrumentalism, prepared to deny the existence of a physical reality\footnote{Or, perhaps more subtly, to deny sense to the \emph{question} of the existence of such a reality.} between quantum measurements. Otherwise, however,  TSAR's ontological prerequisites do not seem especially demanding. Quantum realists will need to look elsewhere for an objection to SRRT. 

\section{Avoiding Discreteness?}

So the best prospect for rejecting SRRT seems to lie with role of Discreteness in TSAR. And on the face of it, a basis for challenge on these grounds is not hard to find. 
As noted above, input-independent control disappears from the QM case, if we allow the Demon the option of introducing a photon in a superposition of the $\textup{\textbf{L}}=1$ and $\textup{\textbf{L}}=0$ cases. In this case, by an appropriate choice of input amplitudes on each channel, the Demon can produce any $\tau_L$ he wishes, whatever Lena's choice of $\sigma_L$. (As in the classical case, this follows immediately from a consideration of the time-reversed case, given the time-symmetry of the relevant quantum dynamics.)

To avoid output-independence, and hence retrocausality, this option needs to be available to Nature in Rena's case.  However, it may seem that Rena can frustrate Nature, simply by placing photon detectors on the output channels. Since such a measurement is guaranteed to yield either $\textup{\textbf{R}}=1$ or $\textup{\textbf{R}}=0$, rather than a superposition of both, it seems that Rena can easily deprive Nature of the flexibility needed to evade the argument for retrocausality. 

But this argument is too quick. In effect, it depends on the assumption that the measurement outcomes constitute the entirety of relevant output ontology  -- in other words, the sole available reservoir, from Nature's point of view, for non-retrocausal variability, in the light of changes Rena may make to $\sigma_R$. Some views of QM will reject this assumption, however.  Everettians will certainly do so, for example. In the Everett picture, there is in reality no single definite outcome, whatever the appearances in any single branch. Rather, the photon emerges from Rena's polarizer in a superposition of  $\textup{\textbf{R}}=1$ and $\textup{\textbf{R}}=0$, with the mod-squared amplitudes for the two components provided by the factors $\cos^{2}(\tau_R - \sigma_R)$ and $\sin^{2}(\tau_R - \sigma_R)$, respectively. Once again, these factors restore the continuous variability of the classical case.

More on this kind of challenge to SRRT in a moment. First, it might be objected that  even without the Everett view (or similar), the option of producing a superposition of $\textup{\textbf{R}}=1$ and $\textup{\textbf{R}}=0$ is still available to Nature; a definite outcome only being needed when Rena makes a measurement. 
 This is true, but I think of limited use at this point, given the constraints of the problem. The task is to explain the correlations observed between inputs and outputs, in a quantum device of the kind depicted in Figure 6; and the issue is whether an explanation can be given using a time-symmetric intermediate dynamics and ontology, without admitting retrocausality. We have noted that in the analogous classical problem in Figure 5, the solution depends on the fact that the outputs may be continuously distributed between $\textup{\textbf{R}}=1$ and $\textup{\textbf{R}}=0$. In the quantum case, it is no help that superposition may provide continuity ``inside the black box,'' so long as the experimental outputs themselves remain discrete.  The Everett view gets off the hook by denying that the experimental outputs really \emph{are} discrete, at the global level. Where there is discreteness in the final conditions of the experiment, however, it doesn't seem to make any difference whether it appears at the time the photon encounters the polarizer, or at some later time.

For the Everett view, the wave function provides Nature with all the flexibility She needs to absorb the consequences of changes in $\sigma_R$, within a time-symmetric ontology, without requiring that they show up at earlier times. And the trick is available to any view that takes an ontological view of the wave function -- so long as it does not throw the advantage away, by allowing the wave function to collapse! In particular, therefore, it is available to the de Broglie-Bohm (dBB) view,\footnote{Or at least to most versions of dBB: in some versions, the wave function may not be ontic, in the required sense.} despite the fact that this theory also provides discrete outputs and inputs. Like Everettians, Bohmians can consistently combine time-symmetric dynamics and ontology with one-way causality, at least so far as TSAR  is concerned.\footnote{I am setting aside the question as to whether the dBB approach can be applied to photons. One justification for this concession is that the argument of \S3.2 could be recast in terms of spin.}

However, to say that this combination is consistent, from a Bohmian point of view, is not yet to recommend it -- even by Bohmian lights. Do Bohmians actually have good reason to think that causality in these experiments is entirely one-way? And if so,  echoing a concern we raised in \S5.3, are there good reasons to break the symmetry in one direction rather than the other? I shall return to this issue in \S7 below.

\subsection{Making the best case for TSAR?}

We have noted that SRRT can be rejected, without resorting to instrumentalism, so long as quantum ontology provides the continuity required to ensure that all changes Rena makes to  $\sigma_R$ can be taken up in the future (thereby avoiding the need for retrocausal changes). The wave function in Everettian and  (most) dBB models provides this continuity in the ontology; and hence TSAR has no direct bite, in these cases.

Addressing the issue from the other end, what is the best way to make a case for TSAR? In my view, it is to begin with a naive operational viewpoint, treating measurement outcomes as real and definite, and as the primary data to be explained; and being at best sceptical concerning claims  that in some far-reaching sense, the ontology of the macroscopic world is anything other than it seems to be. When this starting point is combined with realism -- that is, for these purposes, with the project of trying to understand what \emph{microscopic} reality could account for the observed correlations at the level of measurements -- then TSAR  demonstrates that retrocausality emerges in a natural way from time-symmetry and intuitive causality. The beable structure required for a local causal explanation of correlations between sequential measurements inevitably introduces retrocausality, when made time-symmetric. 

Once again, this case for retrocausality can be avoided by adding an element to the ontology which is not there in the naive picture, and which departs a considerable distance from the naive picture -- a real wave function, surviving measurement. But since it has not previously been appreciated just how directly retrocausality arises from any time-symmetric  naive model, it seems premature, to say the least, to take the fact that TSAR \emph{can} be evaded by these means as a reason for thinking that it \emph{should} be ignored -- and all the more so, because most of the standard reasons for thinking that a naive realist picture won't work in QM (i.e., the No Hidden Variable theorems) have themselves \emph{assumed} that there is no retrocausality. 

This case for not rushing hastily from ``can'' to ``should'' (with respect to the possibility of avoiding the implications of TSAR) might be strengthened if we could challenge the presumption that an Everettian or dBB view does avoid retrocausality. If it does not avoid it, after all, then this way of escaping the implications of TSAR would be a case of out of the frying pan, into the fire, from the point of view of an opponent of retrocausality. 

Indeed, there is a prospect of an interesting dilemma for opponents of retrocausality, if the extra ``non-naive'' ontology needed to block TSAR automatically makes it impossible to exclude retrocausality on observational grounds (by putting it ``out of sight,'' as it were). With this prospect in mind, I now turn to an issue deferred above. What grounds, if any, can an advocate of the dBB view offer that such a theory is \emph{not} retrocausal, in its usual form?\footnote{I set aside for the present the corresponding issue for the Everettian case.}

 \section{Retrocausality in de Broglie-Bohm?}
 
TSAR shows that ``intuitive''  one-way causality must fail, given time-symmetry,  in some versions of a QM account of photon polarization. While the dBB theory is not one of those versions, escaping an argument for the conclusion that the intuitive view is mistaken is not the same as offering an argument that it is not mistaken. It is unclear what form the latter argument might take, for a proponent of the dBB view. To address these issues, we need to step back a bit, and think about what is at stake.  One of the attractions of examples like that of Figure 6 is that they make it easy to pose the relevant questions, without surreptitiously introducing a bias in one direction or other. The symmetries of the model make any bias easily visible.

Intuitive causality clearly involves such a bias: we find it natural to say that if Rena had chosen a different setting, that difference would have made no difference to the photon arriving from the left, but might have affected the result to the right; but very unnatural to say that if Lena had chosen a different setting, that difference would have made no difference to the photon departing to the right, but might have affected the input from the left. We have seen that if the ontology is time-symmetric, and the outputs at $\textup{\textbf{R}}=1$ and $\textup{\textbf{R}}=0$ are discrete in the relevant sense, this asymmetry is unsustainable -- we are forced to revise our view about Rena's case. But where does the intuitive bias come from in the first place? And what reason, if any, do we have in continuing to take it for granted, in a case such as the dBB theory (where abandoning it is not forced on us)?

Our simple model helps with the first of these questions. In the classical case, the obvious difference between Lena and Rena is that Lena controls both the $\sigma_L$ \emph{and} the inputs $\textup{\textbf{L}}=1$ and $\textup{\textbf{L}}=0$; whereas Rena controls only the setting $\sigma_R$. We introduced the Demon in order to restore symmetry, but what breaks symmetry, normally, is the fact that there is no Demon: on the contrary, experimenters often control their inputs, too.\footnote{A fact which relies on the availability of low entropy sources, presumably.} 

But if this is the right story about the source of the bias in the classical case, it seems to cast doubt on our entitlement to retain the bias, in the single-photon case. Intuitively, Lena does have enough control over the classical input beams to ensure that they \emph{would have been the same} (from the same directions, with the same intensities and the same polarizations), if she had chosen a different measurement setting. She does not have enough control over the behaviour of a single-photon source to ensure that \emph{the same photon} would have entered the apparatus, with \emph{the same properties,} if she had chosen a different measurement setting. Intuitive causality tells us that a different setting wouldn't have made any difference; but in this case, unlike in the classical case, the intuition isn't backed up by anything that Lena can actually \emph{do.} 

It might be objected that Lena can do something almost as useful, even in the QM case. She can ensure that no photons enter the apparatus through (say) the $\textup{\textbf{L}}=0$ channel, by simply blocking that channel. Or, if it is objected that this does not exclude thermal photons emitted from the screen itself, she can provide a photon source on the $\textup{\textbf{L}}=1$ of sufficient intensity to guarantee that \emph{most} photons arrive via that channel. 

But if this explains the intuitive causal asymmetry, then the asymmetry should disappear if we deprive Lena of this ability -- e.g., if we assume that the inputs to Lena's apparatus are provided by thermal radiation, with no built-in preference for one channel or the other. Yet  intuition still seems to tell us that Lena's choices still make a difference to the photon between the polarizers, whereas Rena's do not. Is this a mistake? And if not, where does this persistent asymmetry come from?

Furthermore, note that it would not be an option at this point to try to restore symmetry by denying that Lena's choices make a difference to the photon between the polarizers. In that case, the only place to absorb the differences associated with Lena's choices would be in the inputs, and so we would be granting Lena retrocausal influence \emph{over} those inputs. The only way to avoid such retrocausal influence on Lena's part is to insist that her choices make a difference exclusively in the future. But then symmetry would imply that Rena's choices would make a difference exclusively in the past -- so we still face the problem of grounding an \emph{asymmetry,} if we are to avoid retrocausality, in one place or other.

If we wish to insist on this intuitive causal asymmetry, even in the case in which we explicitly exclude the kind of bias introduced by allowing Lena access to low-entropy sources, then there seem to be three possible explanations of its origins:
\begin{enumerate}
\item Some asymmetry in \emph{initial conditions,} other than the usual thermodynamic asymmetry in the sources Lena would normally be taken to control.
\item Some \emph{dynamical} asymmetry in the theory itself.
\item Some \emph{conventional} asymmetry, governing the causal or counterfactual reasoning we use in considering what depends on what.
\end{enumerate}
Let us explore these options in reverse order.

The third option has some plausibility, in my view, as an account of the time-asymmetry of much of the kind of causal and counterfactual reasoning we employ in ordinary life (see, e.g., [\ref{ref:price96}], [\ref{ref:price07}], [\ref{ref:pricewes}]). But it is entirely irrelevant here, where the question is whether there are any good grounds for excluding HV models for QM involving the kind of lawlike correlations between HVs and later measurement settings that have been called ``retrocausal.'' Option (3) might reveal that this is a poor name for the view, but it provides no reason whatsoever to deny the existence of such correlations themselves. In particular, it provides no reason for a proponent of the standard dBB view to the deny the existence of such correlations, \emph{within the standard dBB framework.} (In other words, a proponent of dBB who is convinced that her view excludes such correlations must have something other than Option (3) in mind.)

But Option (2) is off the table, in the case of the standard dBB view, because the dynamics is not time-asymmetric. So the only remaining option is (1) -- i.e., 
 to appeal to some further asymmetry of initial conditions, of a kind not excluded by the elimination of special initial conditions related to the thermodynamic asymmetry. But how are we to justify such conditions? And to show that they are at all in tension with violations of Independence? (It won't do to appeal to the familiar causal asymmetry, if the familiar causal asymmetry relies on something else entirely, namely, the low entropy initial conditions available in typical experiments.) 

 So far as I can see, there is nowhere else for the dBB view to turn at this point, in defence of the claim that the theory is not retrocausal, in its standard form. Once the role of special initial conditions related to the thermodynamic asymmetry is factored out, any  remaining prohibition on retrocausality is merely put in by hand, with no substantial justification whatsoever. Given the distinctive ontology of the dBB view, the required assumption needs to be put in at two points, apparently: it needs to be assumed that \emph{neither} the wave function \emph{nor} the initial positions of the particles are affected by later measurement choices. 
 
 In the absence of any apparent justification for these assumptions, I conclude that while dBB blocks TSAR, it doesn't thereby exclude retrocausality, in the sense of interest in QM. It blocks an argument \emph{from} time-symmetry \emph{to} retrocausality, but this is not the same excluding retrocausality itself.\footnote{For present purposes, I am setting aside  the suggestion  that retrocausality can be excluded on the grounds that it conflicts with free will. See [\ref{ref:price96}], [\ref{ref:evans}], for some critical discussion of this point. In the present context, I don't know any proponent of the dBB view who maintains that it is our knowledge of our own free will that establishes that the dBB theory is entitled to the additional assumptions required to ensure that it is not retrocausal.} 

Intriguingly, there are hints of an argument for the opposite conclusion --  for the view that the dBB theory should \emph{reject} intuitive causality at the fundamental level, in favour of a symmetric picture. Goldstein and Tumulka [\ref{ref:gold03}] offer a dBB ``toy model''  in which microcausal retrocausality provides a Lorentz-invariant explanation of the Bell correlations. This model differs in some ways from other retrocausal proposals (e.g., [\ref{ref:costa}]--[\ref{ref:cramer86}],  [\ref{ref:hokk}], [\ref{ref:miller96}]--[\ref{ref:price96}], [\ref{ref:suth83}]--[\ref{ref:wharton09}]) for reconciling these correlations with special relativity, but the underlying strategy is exactly the same: zig-zag causality, retrocausal on one arm, provides a decomposition of Bell's spacelike correlations into a product of timelike correlations.\footnote{The main difference is that in the proposal in [\ref{ref:gold03}],  the zig-zag goes initially via the future, rather than initially via the past (see [\ref{ref:gold03}], Figure 4, p.~563).} 

Suppose it were to turn out that abandoning intuitive causality at the fundamental level, in favour of a causally symmetric picture, provided a successful route to a Lorentz-invariant formulation of the dBB theory; and that we were convinced that this was the only way to make the theory Lorentz-invariant. We would then have a basis for an analogue of the argument of \S3.2, within the dBB framework, in which Lorentz-invariance played the role of Discreteness: if Nature is constrained to be Lorentz-invariant, She simply does not have the option of absorbing all counterfactual changes of $\sigma_R$  in the future, but must allow Rena a degree of output-independent control of the past.\footnote{For a friend of retrocausality, it would be an enormously satisfying result if two of the great lessons of Einstein's \emph{annus mirabilis,} the quantisation of light and Lorentz-invariance, were to converge in this way, in support of a realist view of the quantum world. }

\subsection{The textbook model, again}

Finally, I would like to note that most of these considerations apply to the textbook model of QM, with the Projection Postulate, as well as to dBB. We observed in \S5.3 that because the textbook model is explicitly time-asymmetric, it evades TSAR. However, as we have just emphasised with respect to dBB, avoiding an argument that entails retrocausality is not the same as establishing that there is no retrocausality -- it is necessary but not sufficient, as it were. 

The discussion in this section seems to apply with almost equal force to the textbook model as it does to dBB. The one difference, I think, is that because the dynamics of the textbook model are not time-symmetric, we cannot immediately exclude option 2 above, as an explanation of the intuitive causal asymmetry. But a little reflection suggests that the dynamical time-asymmetry of the textbook model -- i.e., the Projection Postulate itself -- is of entirely the wrong character for the purpose. To rule out retrocausality, we need to show that changing a measurement setting -- e.g., in our experiment, the value of $\sigma_R$ -- does not change the earlier state of the system measured. The Projection Postulate is simply silent on this matter: it specifies the state \emph{after} a measurement, but says nothing about the state \emph{before} a measurement. As in the case of dBB, then, the textbook model's assumption that the state function is independent of future measurement settings seems simply to be put in by hand, in practice.\footnote{See [\ref{ref:schulman97}] for a retrocausal approach that exploits this fact, by allowing that the initial wave function may indeed depend on what happens to a system in the future.}

\section{Summary}

We began by noting  the claim that time-symmetry favours retrocausality in QM; and the  counter-argument that classical physics seems to permit time-symmetry without retrocausality and that it isn't clear why QM should be different. We have now offered an argument that shows that QM \emph{is} different, under some assumptions about quantum ontology. However, we have also noted that these assumptions are far from compulsory: they are rejected, in one way or another, by several interpretations of QM. 

To put this conclusion in perspective, it is worth stressing  that the relative merits of the views of QM that reject these assumptions -- e.g., instrumentalist views, on one hand, or the Everett  and dBB views, on the other -- depend on the demerits of alternative approaches. In particular, they depend on the issue of the viability of the HV program, in the ``just the particles'' sense -- views which combine the HV program with an epistemic view of the quantum state. 

The present result contributes to clarifying this issue, in the following sense. As we have noted, Bell's Theorem and other No Hidden Variable results rely on the assumption that HVs are independent of future measurement settings --  an assumption with considerable intuitive appeal, against the background of the one-way causality familiar in ordinary life. 

However, the present argument shows that within the framework to which such an HV program is committed -- a framework in which discreteness is \emph{not} offset by non-epistemic continuity at the level of the wave function -- a blanket prohibition on retrocausality is incompatible with time-symmetry, in a manner specific to the QM case. This provides a new reason for re-examining the assumption in question, and for exploring HV approaches that relax it. The photon polarization experiment discussed in  \S3.2  may offer a useful model for investigating these questions, under the assumptions of time-symmetry and discrete outputs.\footnote{See [\ref{ref:evans}] for further relevant discussion of this case, developed by means of an analogy with a photon polarization version of the  EPR-Bohm experiment; and [\ref{ref:price08}] for a more abstract approach to the general project of retrocausal modelling.}

\section*{Acknowledgements}
The ideas in this paper owe a great deal to Ken Wharton, having emerged in discussion about related issues for  our joint paper with Peter Evans [\ref{ref:evans}]. I am also very grateful for comments from Richard Healey, Ruth Kastner, Owen Maroney, David Miller and Rob Spekkens,  from an audience in Cambridge in May 2010, and from an anonymous referee. My research is supported by the Australian Research Council and the University of Sydney. 
\section*{References}

\renewcommand{\theenumi}{\arabic{enumi}}
\renewcommand{\labelenumi}{[\theenumi]}
\begin{enumerate}

\item Aharonov, Y., Bergmann, P.G.~\& Lebowitz, J.L., 1964. ``Time symmetry in the quantum process of measurement'', \textit{Physical Review}, 134, B1410--B1416. [\href{http://prola.aps.org/abstract/PR/v134/i6B/pB1410_1}{DOI: 10.1103/PhysRev.134.B1410}] \label{ref:aharonov}

\item Bell, J.S., 1964. ``On the Einstein Podolsky Rosen Paradox'', \emph{Physics} 1, 195--200. \label{ref:bell}

\item Bell, J.S., 1984. ``Beables for quantum field theory'',  CERN-TH 4035/84, Aug.~2, 1984. Reprinted in \emph{Speakable and Unspeakable in Quantum Mechanics,} 2nd edn., Cambridge: Cambridge University Press, pp.~173--180.\label{ref:bell2}

\item Costa de Beauregard, O., 1953. ``M\'echanique quantique'', \emph{Comptes Rendus  de l'Acad\'emie des Sciences}, T236, 1632--1634. \label{ref:costa}

\item Costa de Beauregard, O.,~1976. ``Time Symmetry and Interpretation of Quantum Mechanics'', \textit{Foundations of Physics},  6, 539--559. \\\hspace{0pt}
[\href{http://www.springerlink.com/content/k8l386556135g250/}{DOI: 10.1007/BF00715107}]\label{ref:costa76}

\item Costa de Beauregard, O., 1977. ``Time symmetry and the Einstein paradox'', \textit{Il Nuovo Cimento}, 42B, 41--63. [\href{http://www.springerlink.com/content/x755748406278g65/}{DOI: 10.1007/BF02906749}]\label{ref:costa77}

\item Cramer, J.G., 1980. ``Generalized absorber theory and the Einstein-Podolsky-Rosen paradox'', \textit{Physical Review D}, 22, 362--376. \\\hspace{0pt}
[\href{http://prd.aps.org/abstract/PRD/v22/i2/p362_1}{DOI: 10.1103/ PhysRevD.22.362}]

\item Cramer, J.G., 1986: ``The transactional interpretation of quantum mechanics'',  \textit{Reviews of Modern Physics}, 58, 647--687. \\\ [\href{http://rmp.aps.org/abstract/RMP/v58/i3/p647_1}{DOI: 10.1103/RevModPhys.58.647}]
 \label{ref:cramer86}

\item Evans, P., Price, H. and Wharton, K.B., 2010. ``New slant on the EPR-Bell experiment'', \emph{British Journal for the Philosophy of Science,} forthcoming. [\href{http://arxiv.org/abs/1001.5057}{arXiv:1001.5057v3 [quant-ph]}] \label{ref:evans}

\item Hokkyo, N., 1988. ``Variational formulation of transactional and related interpretations of quantum mechanics'', \emph{Foundations of Physics Letters,} 1, 293--299. [\href{http://www.springerlink.com/content/p81r939075727216/}{DOI: 10.1007/BF00690070}] \label{ref:hokk}

\item Goldstein, S.~and Tumulka, R., 2003. ``Opposite arrows of time can reconcile relativity and nonlocality'', \emph{Classical and Quantum Gravity,} 20, 557--564. [\href{http://arxiv.org/abs/quant-ph/0105040v5}{arXiv:quant-ph/0105040v5}] [\href{http://www.iop.org/EJ/abstract/0264-9381/20/3/311/}{DOI: 10.1088/0264-9381/20/3/311}]\label{ref:gold03}

\item Miller, D.J., 1996. ``Realism and time symmetry in quantum mechanics'', \emph{Physics Letters,} A222, 31--36. [\href{http://linkinghub.elsevier.com/retrieve/pii/0375960196006202}{DOI: 10.1016/0375-9601(96)00620-2}] \label{ref:miller96}

\item Miller, D.J., 1997. ``Conditional probabilities in quantum mechanics from a time-symmetric formulation'', \emph{Il Nuovo Cimento,} 112B, 1577--1592. 

\item Price, H., 1984. ``The philosophy and physics of affecting the past'', \textit{Synthese}, 16, 299--324. [\href{http://www.springerlink.com/content/p94624888311jv41/}{DOI: 10.1007/BF00485056}]

\item Price, H., 1994. ``A neglected route to realism about quantum mechanics'',  \textit{Mind}, 103, 303--336. [\href{http://arxiv.org/abs/gr-qc/9406028}{arXiv:gr-qc/9406028v1}]

\item Price, H., 1996. \textit{Time's Arrow and Archimedes' Point,} New York: 
Oxford University Press.
\label{ref:price96}

\item Price, H., 1997. ``Time symmetry in microphysics'',  \emph{Philosophy of Science,} 64, S235--244. [\href{http://arxiv.org/abs/quant-ph/9610036}{arXiv:quant-ph/9610036v1}] \label{ref:price97}

\item Price, H., 2007. ``Causal perspectivalism'', in H.~Price and R.~Corry, eds., \emph{Causation, Physics and the Constitution of Reality: Russell's Republic Revisited,} Oxford: Oxford University Press, pp.~250--292.\label{ref:price07}

\item Price, H., 2008. ``Toy models for retrocausality'', \emph{Studies in Hist.~and Phil.~of Mod.~Phys.,} 39, 752--776. [\href{http://arxiv.org/abs/0802.3230}{arXiv:0802.3230v1 [quant-ph]}]\label{ref:price08}

\item Price, H. \& Weslake, B., 2010:  ``The time-asymmetry of causation'',  in H.~Beebee, C.~Hitchcock and P.~Menzies, eds., \emph{The Oxford Handbook of Causation,} Oxford: Oxford University Press, pp.~414--443.\label{ref:pricewes}

\item Schulman, L.S., 1997: \emph{Time's Arrows and Quantum Measurement,}  Cambridge: Cambridge University Press.\label{ref:schulman97}

\item Sutherland, R.I., 1983. ``Bell's theorem and backwards-in-time causality'', \textit{Int.~J.~Th.~Phys.}, 22, 377--384.  [\href{http://www.springerlink.com/content/k8g22512n5w38xu1/}{DOI: 10.1007/BF02082904}]\label{ref:suth83}

\item Sutherland, R.I., 1998. ``Density formalism for quantum theory'', \emph{Foundations of Physics.} 28, 1157--1190. [\href{http://www.springerlink.com/content/u551710rp1x27723/}{DOI: 0.1023/A:1018850120826}]

\item Sutherland, R.I., 2008. ``Causally symmetric Bohm model'', \emph{Studies in Hist.~and Phil.~of Mod.~Phys.,} 39, 782--805. [\href{http://arxiv.org/abs/quant-ph/0601095}{arXiv:quant-ph/0601095v2}] \label{ref:sutherland}

\item Wharton, K.B., 2007. ``Time-symmetric quantum mechanics'', {\em Foundations of  Physics,} 37, 159--168. [\href{http://www.springerlink.com/content/v765h37416011n63/?p=877db523fa624fd582e512cfb9a5bebb&pi=1}{DOI: 10.1007/s10701-006-9089-1}] \label{ref:wharton07}

\item Wharton, K.B., 2009. ``A novel interpretation of the Klein-Gordon equation'', \emph{Foundations of Physics,} 40, 313--332. \\\ [\href{http://www.springerlink.com/content/l14jr4x24pk82521/}{DOI: 10.1007/s10701-009-9398-2}]  [\href{http://arxiv.org/abs/0706.4075}{arXiv:0706.4075v3 [quant-ph]}].
\label{ref:wharton09}

\end{enumerate}
\end{document}